# Nanostructures Design: the Role of Cocatalysts for Hydrogen and Oxygen Generation in Photocatalytic Water Splitting

Zhexu Xi

*Bristol Centre for Functional Nanomaterials, University of Bristol, Bristol, UK*

**Abstract:** Due to the energy supply pressure caused by non-renewable fuels as well as the environment-related issues, the efficient conversion of solar-chemical energy via photo-induced water splitting is one of the promising strategies to address the existing problems. To strengthen the overall catalytic performance of photocatalytic hydrogen ($H_2$) and oxygen ($O_2$) evolution, the selection and construction of cocatalysts are crucial. Recently, semiconductor photocatalysts have been well modified with the loaded cocatalysts as the active sites by extending light harvest, promoting electron separation and transfer, and improving the photocatalytic activity. Combined with the principles of photocatalysis, the paper focuses on the mechanism and roles of cocatalysts for boosted photocatalytic water splitting in recent research. The categories with the corresponding research contents of the existing cocatalysts are also summarised, including cocatalysts for $H_2$ evolution, cocatalysts for $O_2$ evolution, dual cocatalysts for overall water splitting and artificial cocatalyst complexes. Finally, the future direction of the development is suggested for the rational design and large-scale application of highly efficient cocatalysts in the photo-induced water splitting system.

## 1. Introduction

Nowadays, due to rapid growth in population, fast-growing demands for non-renewable, high-pollution energy worldwide have triggered two mainstream issues: one is population and its unbalanced resource allocation; another is environmental problems[1]. In order to address these energy and environmental issues, it's urgent to find and develop the earth-abundant, sustainable and cleaning energy source to replace the traditional fuel energy[2]. Among all of the cleaning energy, solar energy is the richest one on Earth despite its discontinuous spatio-temporal distribution and relatively low energy flow density. Besides, hydrogen energy is always considered as one of the idealist resources due to its sustainability, high energy density and non-toxicity[2][3]. Thus, it's of great value to explore a suitable way of solar-hydrogen energy conversion to give more access to efficient energy storage and utilisation.

Compared with the other common industrial $H_2$ production pathways, solar-driven water splitting has been a constantly appealing research focus because it is a milder and more efficient method[4]. Since the pioneer work of *Fujishima et al.*[5] on $H_2$ and $O_2$ generation via

TiO$_2$-based photocatalytic water splitting shown the boosted activity, more various semiconductors have been developed as heterogeneous photocatalysts to reduce the non-spontaneity in the photo-excited water splitting process. The solar-driven water photolysis mainly contains three steps (**Fig. 1**)[6][7]: (1) light harvesting: when the proton energy surpasses the band gap, the photocatalyst absorbs the light and generates the photo-induced electron-hole (e$^-$-h$^+$) pairs; (2) charge separation and transport: the pairs are separated from each other and migrate to the surface of the catalyst, where the electrons are excited from the valence band (VB) to the conduction band (CB) with the holes left in the VB; (3) surface redox reaction: the electrons in the CB drives the H$_2$ production in the reduction reaction with the VB holes for O$_2$ generation in oxidation. Theoretically, for Step 1, the light absorption ability depends primarily on the band gap of the semiconductor materials; for Step 2 and 3, the corresponding efficiency can be accurately regulated via the coupling effect between catalysts and cocatalysts[8].

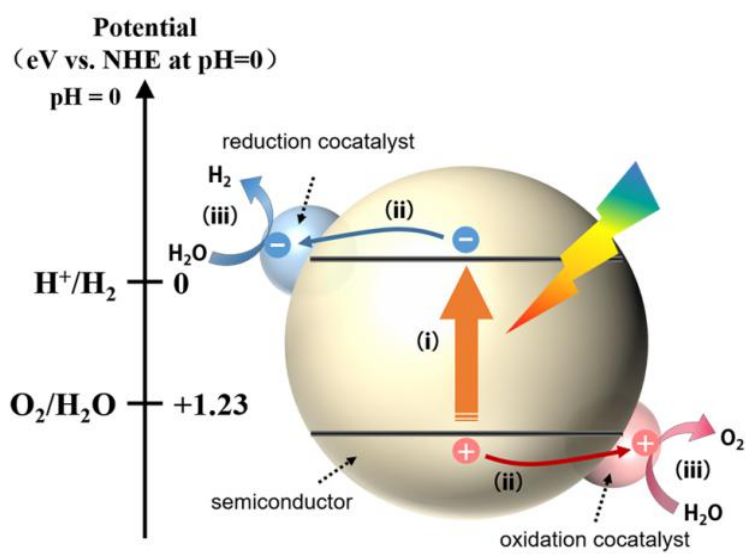

**Fig. 1** Schematic diagram of the mechanism of photocatalytic water splitting containing 2 parts (semiconductor photocatalysts and cocatalysts loaded on the photocatalysts) and 3 steps ((1) light absorption; (2) separation and transfer of charge carriers (electrons and holes); (3) photo-induced hydrogen and oxygen evolution reactions on the surface).[7]

However, at present, numbers of photocatalysts possess insufficient photocatalytic activity for H$_2$/O$_2$ generation because of the following problems: narrow spectral scope, inefficient electron-hole separation, poor surface reaction, and relatively high overpotential[6, 8, 9]. Accordingly, versatile strategies have been implemented to improve photo-excited charge separation and migration for enhanced quantum efficiency, like heteroatom doping, heterojunction construction, morphology modulation, and cocatalysts loaded on the semiconductors[8, 9]. Among the proposed strategies, cocatalysts play an indispensable role in photocatalytic activity and stability. Specifically, cocatalysts can harvest charge carriers for boosted electron-hole separation, expose the abundant active sites for optimised surface redox reactions, lower the overpotential for strengthened photocatalytic activity, and inhibit the light corrosion for stabler photocatalysts.

Herein, we focus on the roles and mechanism of cocatalysts in the photocatalytic water splitting system based on different categories of the existing photocatalysts. We also highlight the difference between a single cocatalyst for $H_2$ or $O_2$ evolution and dual cocatalysts system. Moreover, from the nanoscience and photochemistry perspectives, we also discuss the optimal design of a novel, more efficient and stabler cocatalyst for tuning the surface reaction and charge separation processes.

## 2. The roles of cocatalysts in water photocatalysis

Cocatalysts have been consistently demonstrated to influence positively on the photocatalytic performance and stability of photo-induced water splitting. Loading the proper kind and numbers of cocatalysts on the semiconductor photocatalysts can evidently facilitate the solar conversion efficiency, in spite of the poor intrinsic light absorption ability of semiconductors themselves[6]. Hence, the prominent roles of cocatalysts can be summarised in four aspects, as illustrated in **Fig.2**:

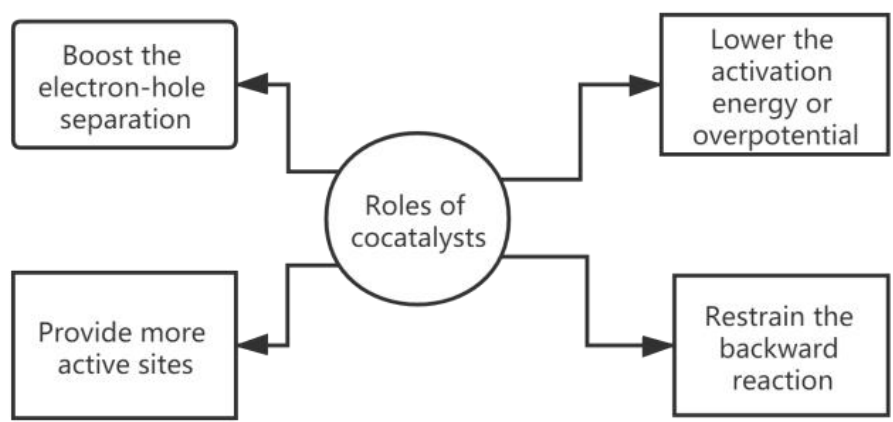

**Fig. 2** Four roles of cocatalysts in photocatalytic $H_2/O_2$ generation via water dissociation on the semiconductor surface

(1) The cocatalysts have more outstanding work function than the semiconductors, so the cocatalysts absorb the photo-generated electrons, thereby facilitating the efficiency in electron-hole separation and migration as well as inhibiting the bulk recombination[10];
(2) By accurately regulating the size, shape, morphology and proportion of the cocatalysts, more catalytically active sites can be exposed to alter the local state of density, thereby enhancing the activity of water splitting[11];
(3) Compared with $H_2$ evolution, the four-electron transfer $O_2$ evolution possesses higher overpotential, so the introduction of single or dual cocatalysts can markedly lower the activation energy (overpotential) of water dissociation and accelerate the $H_2/O_2$ evolution reaction on the surface of photocatalysts[8][12];
(4) The loaded cocatalysts can harvest the photo-excited holes for $O_2$ evolution reaction to avoid the self-decomposition of photocatalysts, thereby increasing the stability of photocatalysts[13].

## 3. Single hydrogen evolution cocatalysts

### 3.1. Noble metals and alloys

On account of the high work function and low Fermi level of noble metals (Pt, Rh, Au, Pd), the Schottky barrier will be generated with the close metal-semiconductor contact. Simultaneously, the existence of the barrier leads to the band bending in the semiconductor at the interface of contact, which facilitates the electron-hole separation and improves the photo-generated electron conductivity. Moreover, noble metals can provide abundant active sites and surface trapping states for accelerated $H_2$ generation[14]. Briefly, noble metals as cocatalysts can extraordinarily improve photocatalytic reactivity. Pt, with the highest work function, is widely developed as the idealist $H_2$ evolution cocatalyst.

Compared with the catalytic performance of the pristine semiconductor photocatalysts, the addition of 1%-3% noble metals as cocatalysts can increase the performance up to several orders of magnitude[15]. Li's endeavour in the role of isolated single Pt atoms sets a typical example in this research field[16]. In 2016, he loaded Pt atoms on g-$C_3N_4$ to facilitate the photocatalytic activity via solid-state deposition. Compared with the poor intrinsic catalytic rate of the bare g-$C_3N_4$, the sample demonstrates the outstanding photocatalytic kinetics with 0.16 wt.% Pt loading through visible-light irradiation. The reason is that the loaded Pt atoms alter the surface state with more active sites exposed and more photo-generated trapping electrons.

Also, the other kinds of noble metals have received public attention based on an entirely different mechanism. Shoaibi et al.[15] loaded Au, Ag, Cu respectively on the surface of $TiO_2$[001] nanocrystals via high-temperature calcination. The boosted photocatalytic activity reveals that the close metal-semiconductor contact strengthens the barrier at the interface and effectively reduces the possibility of electron-hole recombination. More importantly, compared with the other catalytic system, Au/$TiO_2$ shows more extraordinary performance due to the electrical-field-accelerated electron-hole separation induced by the localised surface plasma resonance (LSPR) technique.

Furthermore, the loading of two noble metals together (usually known as alloys or bimetallic cocatalysts) has a synergistic effect and greater tunability in reactivity such as Pt-Pd[17] and Au-Pd[18]. Pt-Pd has been confirmed to be a promising functional cocatalyst for $H_2$ evolution enhancement[17]. The Pt-Pd cocatalyst was deposited on the g-$C_3N_4$ surface to test the different $H_2$ generation rate by regulating the atomic arrangements. Accordingly, the sample in the formula of $Pt_{0.5}Pd_{0.5}$/g-$C_3N_4$ depicts the optimum photocatalytic activity with the gas production rate of 2885.0 µmol·g$^{-1}$·h$^{-1}$.

### 3.2. Transition metals and alloys

Although noble metals and their bimetallic alloys exhibit superior photocatalytic activity, their high costs and scarce reserves severely restrict the large-scale application. Cheaper transition metals (Ni, Cu, Cd, etc.) and their alloys have been regarded as a suitable alternative to noble metals with comparatively great activity and similar catalytic mechanisms linked intimately with the existence of the Schottky barriers.

Sun et al.[19] reported an ultrathin Ni/CdS nanosheets to realise the coupling synthesis of biomass intermediates and $H_2$ generation with the evidently boosted activity. Here, the Fermi level of Ni is just between the bottom of the semiconductor's CB and the $H_2O/H_2$ redox potential as well as the high work function of Ni, so the charge separation is accelerated to improve the quantum efficiency of $H_2$ production. Similarly, the characterisation and photochemical measurement of NiCd/CdS nanorods revealed a synergistic enhancement in photocatalytic activity and shown a drastically boosted $H_2$ production rate of 11.57 mmol·$g^{-1}$·$h^{-1}$, which is approximately 64.8, 17.2 and 2.3 times rate of pristine CdS, Cd/CdS and Ni/CdS, respectively[20]. By partially replacing Ni atoms with Cd atoms, the formation of the NiCd alloy ensures the high work function and strong adsorption energy of the H atom.

### 3.3. Metal oxides and hydroxides

Metal oxides and hydroxides are also commonly used as highly efficient cocatalysts to strengthen the $H_2$ photocatalytic activity. On one hand, by lessening the overpotential for $H_2$ generation, the water-splitting process can be markedly activated with cocatalysts. On the other hand, these kinds of cocatalysts help optimise the surface hole-trapping state to extract more holes and accelerate the charge separation for water photocatalysis enhancement.

Both noble-metal oxides like $TiO_2$[21] and transition-metal compounds like $Cu_2O$[22], NiO[23] and $Ni(OH)_2$[24] possess the proper standard potentials ($E^0(MOx/M)$ or $E^0(M(OH)x/M)$) which are located between the bottom of the CB of the semiconductor and the standard reduction potential for $H^+/H_2$ (0 eV). Consequently, the energy barrier for photo-induced interband transition is decreased and the electron-hole separation is facilitated. Lee et al.[22] further explored the size-dependent confinement in the activity and kinetics of the photocatalytic water splitting. As the size of $Cu_2O$ increases, the energy gap will be reduced to alter the edge of CB and VB for more significantly improved electron transfer, thereby optimising the $H_2$ generation.

### 3.4. Metal sulfides

Transitional-metal sulfides do not only have a huge surface area with plentiful active sites but also possess tailored properties due to various electronic states and well-tuned band structures. As a sort of prospect $H_2$ evolution cocatalyst, they can precisely tune the electronic structures to optimise the photocatalytic activity in three ways[6][25]: (1) boosting the charge separation and transfer; (2) exposing active sites for water dissociation activation; (3) increasing more trapping electrons to retard the recombination.

Yin's work[26] reported a $MoS_2$/CdS heterojunction formed by the partially crystallised $MoS_2$ nanosheets incorporated into the CdS nanorods. This vertically arrayed nanostructure exposes abundant active sites for efficiently extracting electrons, thereby prompting the transport of charge carriers. Simultaneously, the junction reduces the distance of electron transport and largely relieves the electron-hole recombination. Thus, the rate of photo-induced $H_2$ production is boosted.

### 3.5. Metal phosphides

Transitional-metal phosphides are universally studied due to their excellent activity, high stability and earth-abundant reserves. Moreover, the ensemble effect of P-ligand forms a moderate binding strength to H atoms, thereby contributing to the surface adsorption and desorption kinetics when dissociating water and prompting the $H_2$ production rate[27].

One of the representative work related to metal phosphides as cocatalysts in the water photolysis system is verifying the photocatalytic scalability of $Ni_2P$-decorated $TiO_2$, $g-C_3N_4$ and CdS[28]. The result showed that the decoration of $Ni_2P$ could enhance the transfer of charge carriers and optimise the surface characteristics for the improved $H_2$ photo-reduction rate and outstanding quantum efficiency.

## 4. Single oxygen evolution cocatalysts

Compared with the $H_2$ evolution reaction, $O_2$ evolution, entailing the sluggish four-electron transfer kinetics with the high activation energy needed, is always regarded as the major bottleneck of overall water splitting[29]. Ideal oxygen evolution cocatalysts can act effectively as a hole mediator to decrease the overpotential of $O_2$ generation and facilitate the corresponding kinetics and activity. Besides, the high robustness and catalytic selectivity are also important indexes to evaluate the quality of an ideal cocatalyst.

### 4.1. Metal oxides and hydroxides

Noble-metal oxides and transition-metal oxides and hydroxides are most broadly investigated for enhanced $O_2$ evolution. For noble-metal oxides, $IrO_2$ and $RuO_2$ are two of the typically efficient cocatalysts[30]. Ir-based and Ru-based oxides exhibit superior photocatalytic performance in most $O_2$ evolution reactions (OER), but several factors restrict their wide-ranging application, like the high cost and rare elemental reserves of $IrO_2$ and the instability and short service lifetime of $RuO_2$[31]. To solve these problems, Nguyen et al.[30] synthesised size-controllable $IrO_2$ and $RuO_2$ nanoparticles for OER enhancement, where the size was precisely tuned by altering the annealing temperature. Here, $RuO_2$ presents an extremely high mass activity for boosted OER and $IrO_2$ exhibits a remarkable specificity of $O_2$ generation.

A series of transition-metal (Co, Ni, Mn, Fe, etc.) oxides and hydroxides are considered promising to replace noble-metal oxides owing to their comparatively great photocatalytic activity, strong light corrosion resistance and low cost. Co-based compounds can abate the energy barrier for $O_2$ evolution activation and generate an internal electrical field with semiconductor catalysts for strengthened charge transfer[32]. According to Wang's work[33], the rational design of $g-C_3N_4$-supported $Co_3O_4$ quantum dots as the cocatalyst, which is mainly based on the precise regulation of the loading capacity of the cocatalyst and the calcination temperature of the sample, promotes the efficiency of electron-hole separation and migration and results in an evidently enhanced OER activity. Wang et al.[34] deposited Co(OH)$_2$ on g-C$_3$N$_4$, facilitating the charge transport as well as reducing the energy of O-O bond formation for boosted O$_2$ generation.

### 4.2. Bimetallic oxides

The bimetallic system shows a superior synergistic effect, better tunability and selectivity in modulating the electronic structure of the cocatalyst and enhancing the water activation compared with the single metal-based system.

Wu et al.[35] synthesised a high-activity cocatalyst, amorphous FeMnOS, to boost the O$_2$ photo-oxidation rate of BiVO$_4$ as a photoanode material. The synergistic effect of Fe/Sn in the nanostructures markedly lowers the overpotential and accelerate the surface OER kinetics with a 3.4 times higher photocurrent density than that of the pristine BiVO$_4$ photoanode and its onset potential shifting negatively from 0.44 V to 0.25 V.

### 4.3. Metal phosphides and phosphates

Transition-metal phosphides and phosphates serve as effective hole mediators to promote the surface charge transport for boosted O$_2$ production activity[8][36]. One representative research primarily focuses on the correlations between the OER performance and the loading amount of layered cobalt phosphate (CoPi) complexes through coordination tuning[37]. Besides the enhanced catalytic reactivity, the workers also found that as the percentage of loaded CoPi increased, the OER activity changed in a volcano-like trend. To explain more specifically, the moderate loading amount can be beneficial to the charge separation and surface reaction for water splitting enhancement; when the loading amount surpasses the optimum value, the excess cocatalyst will cover the photocatalyst surface to hinder the water activation and dissociation.

## 5. Dual cocatalysts for overall water splitting

High-efficiency photochemical water splitting is closely related to both reduction and oxidation half reactions. Although the above research has fully demonstrated the superior photocatalytic performance with the incorporation of the cocatalyst in the single H$_2$ or O$_2$ evolution system, the dual cocatalysts can act more efficiently in facilitating the electron reduction and hole oxidation simultaneously. Specifically, the reductive cocatalysts act as electron trapping agents to help reduce the proton to H$_2$, while the oxidative cocatalysts tend to extract holes for the oxidation reaction. Consequently, the coupling coordination of the dual cocatalysts can effectively eliminate the fast recombination of charge carriers, accelerate the surface charge transport and improve the overall water splitting efficiency[38][39]. What is more, a robust photocatalytic system with dual cocatalysts loading requires a stoichiometric ratio of its products (H$_2$:O$_2$) to be 2:1. Thirdly, regarding dual cocatalysts, the rational modulation of their content, size, morphology and distribution can optimise the spatial separation and migration of charge carriers and avoid the backward reaction (H$_2$+O$_2$→H$_2$O). For example, Demon et al.[39] completed the overall water photocatalysis with the RuO$_x$/Cr$_2$O$_3$ complex as the reductive cocatalyst and IrO$_2$ as the oxidative cocatalyst, which shown an evidently boosted quantum efficiency of overall water splitting compared with that in the pure $RuO_x/Cr_2O_3$ system. Interestingly, the amorphous $Cr_2O_3$ shell is tested to be oxygen-permselective, so this structure effectively eradicates the $H_2$-$O_2$ recombination.

For the simultaneous enhancement of $H_2$ and $O_2$ generation, feasible strategies in modulating the band structures and electronic states of cocatalysts are essential. These strategies are beneficial for extracting electrons and holes and stimulating spatial charge separation. As a kind of ideal 2D-layered material, $C_3N_4$ has a well-tuned morphology and electronic structure due to its planar, conjugated structure, which easily gets constant attention from Wang's group[40-42]. The first example of their research is considered the first to realise the efficient water splitting based on a polymeric semiconductor system ($Pt/CoO_x/C_3N_4$), which possesses stable production rates of 1.2 μmol/h in $H_2$ and 0.6 μmol/h in $O_2$[40]. Herein, Pt acts in a mixed-valence form: the elementary Pt acts for $H_2$ evolution enhancement; the oxidation state acts as an $O_2$-generation active site. Next, they incorporated CoP as an efficient oxidative cocatalyst into the water photolysis system. Accordingly, the energy barrier of the oxidation half reaction was declined for promoted $O_2$-generation activity; additionally, the photo-generated electron-hole separation and migration process was facilitated by the synergistic effect of Pt/CoP cocatalysts. As **Fig. 3** illustrates, the photocatalytic activity and stability are both enhanced compared with those of pristine Pt and CoP water splitting system[41]. After that, they successfully demonstrated how surface redox water splitting is prompted by separating incompatible oxidation and reduction sites using a hollow nanostructure with a permeable semiconductor shell. Specifically, they used mesoporous-$SiO_2$/Pt/$SiO_2$ as the template to load the Pt and $Co_3O_4$ nanoparticles onto the inner and outer shell surface of HCNS, respectively, yielding the $Co_3O_4$/HCNS/Pt system. In **Fig. 4**, the photocatalytic gas evolution rate in this system is nearly three-times faster than that of the random distribution of Pt and $Co_3O_4$ nanoparticles on the HCNS outer surface with the same loading amount (($Co_3O_4$+Pt)/HCNS). This research accomplishes the high-selectivity spatial separation of photo-excited oxidation and reduction active sites via orientated distribution of Pt and Co-based oxides, showing a superior synergistic effect in apparent water splitting[42].

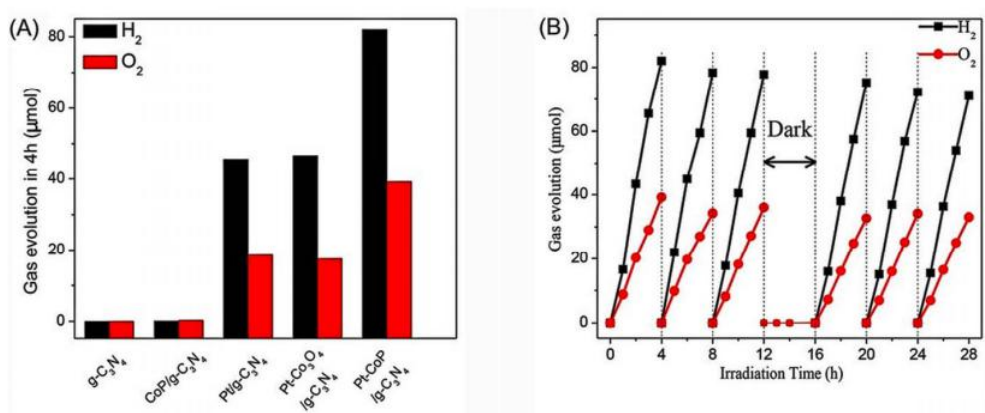

**Fig. 3 (A)** Photocatalytic overall water splitting using $C_3N_4$ loaded with different cocatalysts under full arc irradiation of a Xe lamp at pH = 3 for 4 h, and **(B)** overall water splitting by Pt (3%)-CoP (3%)/$C_3N_4$[41].

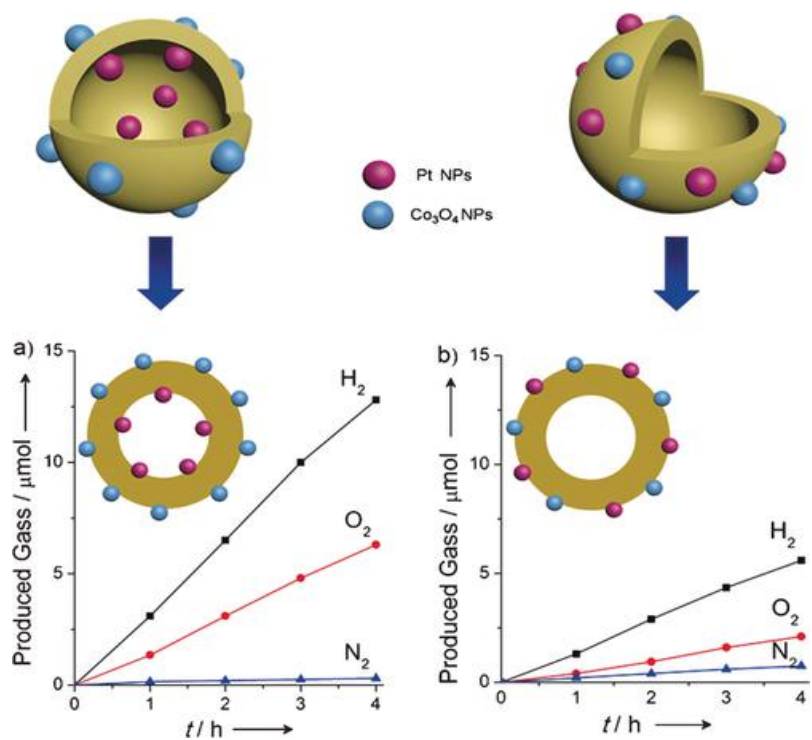

**Fig. 4** Comparison of time courses of photocatalytic evolution of H2 and O2 using **a)** $Co_3O_4$/HCNS/Pt and **b)** ($Co_3O_4$+Pt)/HCNS under UV irradiation (λ>300 nm) in Wang's group[42].

Furthermore, photocatalytic performance can be accurately regulated by altering the cocatalyst-semiconductor contact because the randomly distributed reductive and oxidative cocatalysts will lead to the inevitable random flow of photo-generated charges and fast recombination. Hence, the rational design of dual cocatalysts contributes to the enhanced activity of overall water splitting. Like orderly distributed Pt and $Co_3O_4$ nanoparticles on the inner and outer HCNS shell surface, the spatial separation of reductive and oxidative sites demonstrates the irreversible charge migration, thereby decreasing the possibility of charge recombination[42]. Another common way is constructing 0D/2D complexes. This system exposes more active sites and electron traps and ensures the intrinsic light absorption properties for promoted water splitting performance, like the Pt/$C_3N_4$[40] and the Pt/CoP/$C_3N_4$[41] photocatalytic system mentioned above.

As a result, these examples perfectly exhibits the practical gas generation performance of the ternary photocatalytic system (including semiconducting photocatalysts for light harvesting, oxidative cocatalysts for capturing holes and reductive cocatalysts for trapping electrons), which results in the effective spatial separation of electron-hole pairs, promotes the surface reaction and facilitates the photo-generated water splitting.

## 6. Cocatalyst complexes for activated water adsorption and dissociation

The common dual cocatalysts are spatially separated, so it is hard to realise the water activation simultaneously in the reduction and oxidation half-reactions. Accordingly, the highly activated water molecules are useful in accelerating the surface kinetics of water dissociation[6][43].

Zou et al.[43] designed a novel Pt/Ni(OH)$_2$ cocatalyst with a strong synergistic effect. The results of their calculation and experiments highlight the intense Pt$^{\delta+}$-O$^{\delta-}$-Ni$^{\delta+}$ interaction on the intimate contact of the Pt-Ni(OH)$_2$ interface. Besides consolidating the electron transfer, the interaction also strengthens the water adsorption ability during the surface reaction, making the H-OH bond easier to activate and thereby lessening the energy barrier of water dissociation. Loaded on the C$_3$N$_4$, the Pt-Ni(OH)$_2$ composite cocatalyst shows the maximal apparent quantum efficiency of 1.8% under 420 nm light irradiation; more interestingly, loaded on the anatase (TiO$_2$), the cocatalyst realises the overall water splitting, which cannot be realised in the regular dual cocatalyst system.

## 7. Conclusion and outlook

Energy crisis and environmental issues have always become a long-term focus worldwide, so semiconductor-based photocatalytic water splitting is regarded as an efficient approach for solar-to-chemical energy conversion. In order to improve the H$_2$ and O$_2$ evolution rate, the loading cocatalyst is usually needed in a highly active and robust photocatalyst. The cocatalyst can effectively enhance the light harvest property, prompt the photo-induced separation and migration of charge carriers, expose more sites for gas generation, lower the energy barrier of water activation, and suppress the backward reaction (recombination). The rational design and modulation of cocatalysts via the close contact of semiconductor photocatalysts have been demonstrated to enhance the photocatalytic activity and stability significantly. Here, we summarise the progress of the state-of-the-art cocatalysts over various categories, including single hydrogen generation (reductive) cocatalysts, single oxygen generation (oxidative) cocatalysts, dual cocatalysts and cocatalyst complexes.

Currently, the cocatalysts in photocatalysis have evolved from noble metals and alloys to transition metals and alloys, from single to complex systems. Ulteriorly, the cocatalyst complexes can activate the water molecules for the strengthened kinetics of surface charge transportation. Hence, the structure of cocatalysts and the relationship between catalysts and cocatalysts need further exploration based on the future development of the cocatalyst.
(1) The construction of the multi-component and multi-functional cocatalysts will be crucial to accelerate the surface reactions for promoted overall efficiency;
(2) The parameters of cocatalysts, including the morphology, structures, defects and coordination status, will be better regulated based on the further development of the related theories;
(3) In situ tracking and characterisation system should be further developed to for a more comprehensive understanding of the internal mechanisms of photo-induced water splitting and hydrogen and oxygen generation.

## References

[1] Wu, M.; Zhang, J.; He, B. -B.; Wang, H. -W.; Wang, R.; Gong, Y.-S. In-Situ Construction of Coral-Like Porous P-Doped g-C$_3$N$_4$ Tubes with Hybrid 1D/2D Architecture and High Efficient Photocatalytic Hydrogen Evolution. *Appl. Catal. B: Environ.* **2019**, *241*, 159-166.

[2] Maeda, K. Photocatalytic Water Splitting Using Semiconductor Particles: History and Recent Developments. *J. Photochem. Photobio. C* **2011**, *12*, 237-268.

[3] Turner, J. A. A Realizable Renewable Energy Future. *Science* **1999**, *285*, 687-689.

[4] Fan, K.; Jin, Z. L.; Yang, H.; Liu, D. D.; Hu, H. Y.; Bi, Y. P. Promotion of the Excited Electron Transfer over Ni- and Co -Sulfide Co-doped g-C$_3$N$_4$ Photocatalyst (g-C$_3$N$_4$/Ni$_x$Co$_{1-x}$S$_2$) for Hydrogen Production under Visible Light Irradiation. *Sci. Rep.* **2017**, *7*, 7710.

[5] Fujishima, A.; Honda, K. Electrochemical Photolysis of Water at a Semiconductor Electrode. *Nature* **1972**, *238*, 37-38.

[6] Hisatomi, T.; Kubota, J.; Domen, K. Recent Advances in Semiconductors for Photocatalytic and Photoelectrochemical Water Splitting. *Chem. Soc. Rev.* **2014**, *43*, 7520-7535.

[7] Maeda, K.; Teramura, K.; Lu, D.; Takata, T.; Saito, N.; Inoue, Y.; Domen, K. Photocatalyst Releasing Hydrogen from Water-Enhancing Catalytic Performance Holds Promise for Hydrogen Production by Water Splitting in Sunlight. *Nature* **2006**, *440*, 295.

[8] Sun, S.; Zhang, X.; Liu, X.; Pan, L.; Zhang, X.; Zou, J. Design and Construction of Cocatalysts for Photocatalytic Water Splitting. *Acta Phys. -Chim. Sin.* **2020**, *36*, 1905007.

[9] Moriz, S. A.; Shevlin, S.; Martin, D.; Guo, Z. -X.; Tang, J. Visible-Light Driven Heterojunction Photocatalysts for Water Splitting-A Critical Review. *Energy Environ. Sci.* **2015**, *8*, 731-759.

[10] Bi, W.; Li, X.; Zhang, L.; Jin, T.; Zhang, L.; Zhang, Q.; Luo, Y.; Wu, C.; Xie, Y. Molecular Co-catalyst Accelerating Hole Transfer for Enhanced Photocatalytic H$_2$ Evolution. *Nat. Commun.* **2015**, *6*, 8647.

[11] Chen, H.; Jiang, D.; Sun, Z.; Irfan, R. M.; Zhang, L.; Du, P. Cobalt Nitride as an Efficient Cocatalyst on CdS Nanorods for Enhanced Photocatalytic Hydrogen Production in Water. *Catal. Sci. Technol.* **2017**, *7*, 1515-1522.

[12] Zhang, G.; Lan, Z. -A.; Wang, X. Surface Engineering of Graphitic Carbon Nitride Polymers with Cocatalysts for Photocatalytic Overall Water Splitting. *Chem. Sci.* **2017**, *8*, 5261-5274.

[13] Meng, A.; Zhang, L.; Cheng, B.; Yu, J. Dual Cocatalysts in TiO$_2$ Photocatalysis. *Adv. Mater.* **2019**, *31*, 1807660.

[14] Leung, D.; Fu, X.; Wang, C.; Ni, M.; Leung, M.; Wang, X.; Fu, X. Hydrogen Production over Titania-Based Photocatalysts. *Chemsuschem* **2010**, *3*, 681-694.

[15] Shoaib, A.; Ji, M.; Qian, H.; Liu, J.; Xu, M.; Zhang, J. Noble Metal Nanoclusters and Their In Situ, Calcination to Nanocrystals: Precise Control of Their Size and Interface with TiO$_2$, Nanosheets and Their Versatile Catalysis Application. *Nano Research* **2016**, *9*, 1763-1774.

[16] Li, X.; Bi, W.; Zhang, L.; Xie, Y. Single-Atom Pt as Co-catalyst for Enhanced Photocatalytic H$_2$ Evolution. *Adv. Mater.* **2016**, *28*, 2427-2431.

[17] Xiao, N.; Li, S.; Liu, S.; Xu, B.; Li, Y.; Gao, Y.; Ge, L.; Lu, G. Novel PtPd Alloy Nanoparticle-Decorated g-C$_3$N$_4$ Nanosheets with Enhanced Photocatalytic Activity for H$_2$ Evolution under Visible Light Irradiation. *Chin. J. Catal.* **2019**, *40*, 352-361.

[18] Han, C.; Gao, Y.; Liu, S.; Ge, L.; Xiao, N.; Dai, D.; Xu, B.; Chen, C. In-Situ Synthesis of Novel

Plate-Like Co(OH)$_2$ Co-catalyst Decorated TiO$_2$ Nanosheets with Efficient Photocatalytic H$_2$ Evolution Activity. *Int. J. Hydrogen Energy* **2017**, *42*, 22765-22775.

[19] Han, G.; Jin, Y. -H.; Burgess, R.; Dickenson, N.; Cao, X. -M.; Sun, Y. Visible-Light-Driven Valorization of Biomass Intermediates Integrated with H$_2$ Production Catalyzed by Ultrathin Ni/CdS Nanosheets. *J. Am. Chem. Soc.* **2017**, *139*, 15584-15587.

[20] Wang, B.; He, S.; Zhang, L.; Huang, X.; Gao, F.; Feng, W.; Liu, P. CdS Nanorods Decorated with Inexpensive NiCd Bimetallic Nanoparticles as Efficient Photocatalysts for Visible-Light-Driven Photocatalytic Hydrogen Evolution. *Appl. Catal. B: Environ.* **2019**, *243*, 229-235.

[21] Tan, Y.; Shu, Z.; Zhou, J.; Li, T.; Wang, W.; Zhao, Z. One-Step Synthesis of Nanostructured g-C$_3$N$_4$/TiO$_2$ Composite for Highly Enhanced Visible-Light Photocatalytic H$_2$ Evolution *Appl. Catal. B: Environ.* **2018**, *230*, 260-268.

[22] Karthikeyan, S.; Kumar, S.; Durndell, L.; Isaacs, M.; Parlett, C.; Coulson, B.; Douthwaite, R.; Jiang, Z.; Wilson, K.; Lee, A. Size-Dependent Visible Light Photocatalytic Performance of Cu$_2$O Nanocubes. *ChemCatChem* **2018**, *10*, 3554-3563.

[23] Li, L.; Cheng, B.; Wang, Y.; Yu, J. Enhanced Photocatalytic H$_2$-Production Activity of Bicomponent NiO/TiO$_2$ Composite Nanofibers. *J. Colloid Interf. Sci.* **2015**, *449*, 115-121.

[24] Mao, L.; Ba, Q.; Jia, X.; Liu, S.; Liu, H.; Zhang, J.; Li, X.; Chen, W. Ultrathin Ni(OH)$_2$ Nanosheets: A New Strategy for Cocatalyst Design on CdS Surfaces for Photocatalytic Hydrogen Generation. *RSC Adv.* **2019**, *9*, 1260-1269.

[25] Han, B.; Liu, S.; Zhang,N.; Xu. Y. -J.; Tang. Z. -R. One Dimensional CdS@MoS$_2$ Core-Shell Nanowires for Boosted Photocatalytic Hydrogen Evolution under Visible Light. *Appl. Catal. B-Environ.* **2016**, *202*, 298-304.

[26] Yin, X. -L.; Li, L. -L.; Jiang, W. -J.; Zhang, Y.; Zhang, X.; Wan, L. -J.; Hu, J. -S. MoS$_2$/CdS Nanosheets-On-Nanorod Heterostructure for Highly Efficient Photocatalytic H$_2$ Generation under Visible Light Irradiation. *ACS Appl. Mater. Interfaces* **2016**, *8*, 15258-15266.

[27] Pei, Y.; Cheng, Y.; Chen, J.; Smith, W.; Dong, P.; Ajayan, P. M.; Ye, M.; Shen, J. Recent Developments of Transition Metal Phosphides as Catalysts in the Energy Conversion Field. *J. Mater. Chem. A* **2018**, *6*, 23220-23243.

[28] Chen, Y.; Qin, Z. General Applicability of Nanocrystalline Ni$_2$P as a Noble Metal Free Cocatalyst to Boost Photocatalytic Hydrogen Generation. *Catal. Sci. Technol.* **2016**, *6*, 8212-8221.

[29] Chen, S.; Takata, T.; Domen, K. Particulate Photocatalysts for Overall Water Splitting. *Nat. Rev. Mater.* **2017**, *2*, 17050.

[30] Nguyen, T.; Scherer, G.; Xu, Z. A Facile Synthesis of Size-Controllable IrO$_2$ and RuO$_2$ Nanoparticles for the Oxygen Evolution Reaction. *Electrocatalysis* **2016**, *7*, 420-427.

[31] Liu, B.; Wang, S.; Wang, C.; Chen, Y.; Ma, B.; Zhang, J. Surface Morphology and Electrochemical Properties of RuO$_2$-Doped Ti/IrO$_2$-ZrO$_2$ Anodes for Oxygen Evolution Reaction. *J. Alloys Compd.* **2019**, *778*, 593-602.

[32] Li, M.; Bai, L.; Wu, S. J.; Wen, X. D.; Guan, J. Q. Co/CoO$_x$ Nanoparticles Embedded on Carbon for Efficient Catalysis of Oxygen Evolution and Oxygen Reduction Reactions. *ChemSusChem* **2018**, *11*, 1722-1727.

[33] Zhang, H. Y.; Tian, W. J.; Zhou, L.; Sun, H. Q.; Tade, M.; Wang, S. B. Monodisperse Co$_3$O$_4$ Quantum Dots on Porous Carbon Nitride Nanosheets for Enhanced Visible-Light-Driven Water Oxidation. *Appl. Catal. B: Environ.* **2017**, *223*, 2-9.

[34] Zhang, G. G.; Zang, S. H.; Wang, X. C. Layered Co(OH)$_2$ Deposited Polymeric Carbon Nitrides for Photocatalytic Water Oxidation. *ACS Catal.* **2015**, *5*, 941-947.

[35] Hu, R.; Meng, L.; Zhang, J.; Wang, X.; Wu, S.; Wu, Z.; Zhou, R.; Li, L.; Li, D. -S.; Wu, T. A High-Activity Bimetallic OER Cocatalyst for Efficient Photoelectrochemical Water Splitting of BiVO$_4$. *Nanoscale* **2020**, *12*, 8875-8882.

[36] Kanan, M.; Nocera, D. In Situ Formation of an Oxygen-Evolving Catalyst in Neutral Water Containing Phosphate and Co$^{2+}$. *Science* **2008**, *321*, 1072-1075.

[37] Zeng, F.; Li, J.; Hofmann, J.; Bisswanger, T.; Stampfer, C.; Hartmann, H.; Besmehn, A.; Palkovitsa, S.; Palkovits, R. Phosphate-Assisted Efficient Oxygen Evolution Over Finely Dispersed Cobalt Particles Supported on Graphene. *Catal. Sci. Technol.* **2021**, *11*, 1039-1048.

[38] Dong, J.; Shi, Y.; Huang, C.; Wu, Q.; Zeng, T.; Yao, W. A New and Stable Mo-Mo$_2$C Modified g-C$_3$N$_4$ Photocatalyst for Efficient Visible Light Photocatalytic H$_2$ Production. *Appl. Catal. B: Environ.* **2019**, *243*, 27-35.

[39] Maeda, K.; Lu, D. L.; Domen, K. Direct Water Splitting into Hydrogen and Oxygen under Visible Light by using Modified TaON Photocatalysts with d$^0$ Electronic Configuration. *Chemistry* **2013**, *19*, 4986-4991.

[40] Zhang, G.; Lan, Z.; Lin, L.; Lin, S.; Wang, X. Overall Water Splitting by Pt/g-C$_3$N$_4$ Photocatalysts without Using Sacrificial Agents. *Chem. Sci.* **2016**, *7*, 3062-3066.

[41] Pan, Z.; Zheng, Y.; Guo, F.; Niu, P.; Wang, X. Decorating CoP and Pt Nanoparticles on Graphitic Carbon Nitride Nanosheets to Promote Overall Water Splitting by Conjugated Polymers. *ChemSusChem* **2017**, *10*, 87-90.

[42] Zheng, D.; Cao, X. -N.; Wang, X. Precise Formation of a Hollow Carbon Nitride Structure with a Janus Surface to Promote Water Splitting by Photoredox Catalysis. *Angew. Chem. Int. Ed.* **2016**, *55*, 11512-11516.

[43] Sun, S.; Zhang, Y. -C.; Shen, G.; Wang, Y.; Liu, X.; Duan, Z.; Pan, L.; Zhang, X.; Zou, J. -J. Photoinduced Composite of Pt Decorated Ni(OH)$_2$ as Strongly Synergetic Cocatalyst to Boost H$_2$O Activation for Photocatalytic Overall Water Splitting. *Appl. Catal. B: Environ.* **2019**, *243*, 253-261.